\documentclass[usenatbib]{mnras}

\usepackage{graphicx}

\usepackage{amsmath}
\usepackage{amssymb}
\DeclareMathOperator{\sinc}{sinc}

\title{Cosmological constraints from the convergence 1-point probability distribution}

\author[K. Patton et al.]{
Kenneth Patton,$^{1}$
Jonathan Blazek,$^{1,2}$
Klaus Honscheid,$^{1}$
Eric Huff,$^{1,3}$
\newauthor
Peter Melchior,$^{4}$
Ashley J. Ross,$^{1}$
and Eric Suchyta$^{1,5}$
\\
$^{1}$Center for Cosmology and AstroParticle Physics, The Ohio State University, 191 W. Woodruff Ave., Columbus, OH 43210, USA\\
$^{2}$Laboratoire d’Astrophysique, Ecole Polytechnique Fédérale de Lausanne (EPFL), Observatoire de Sauverny, CH-1290 Versoix, Switzerland\\
$^{3}$Jet Propulsion Laboratory, California Institute of Technology, 4800 Oak Grove Dr., Pasadena, CA 91109, USA\\
$^{4}$Department of Astrophysical Sciences, Princeton University, Princeton, NJ 08544, USA\\
$^{5}$Computer Science and Mathematics Division, Oak Ridge National Laboratory, Oak Ridge, TN 37831, USA\\
}

\pubyear{2016}

\begin{document}

\label{firstpage}
\maketitle

\begin{abstract}
We examine the cosmological information available from the 1-point probability distribution (PDF) of the weak-lensing convergence field, utilizing fast {\sc L-PICOLA} simulations and a Fisher analysis. We find competitive constraints in the $\Omega_m$-$\sigma_8$ plane from the convergence PDF with $188\ arcmin^2$ pixels compared to the cosmic shear power spectrum with an equivalent number of modes ($\ell < 886$). The convergence PDF also partially breaks the degeneracy cosmic shear exhibits in that parameter space. A joint analysis of the convergence PDF and shear 2-point function also reduces the impact of shape measurement systematics, to which the PDF is less susceptible, and improves the total figure of merit by a factor of $2-3$, depending on the level of systematics. Finally, we present a correction factor necessary for calculating the unbiased Fisher information from finite differences using a limited number of cosmological simulations.
\end{abstract}

\section{Introduction}
\label{sec:Introduction}

Weak gravitational lensing provides a powerful probe of both geometry and structure growth in the Universe, probing the underlying distribution of dark matter through correlations in the observed shapes of background galaxies (e.g. \citealt{Bartelmann01} for review). Numerous works have examined the information on the cosmological density field encoded by the distribution of galaxy shapes; these works have primarily made use of the fields' 2-point statistics (known as `cosmic shear'; e.g. \citealt{Abbott16} and references there-in). The 2-point information in the power spectrum (or, equivalently, the correlation function) contains the complete statistical characterization for Gaussian fields. For an initially Gaussian density field, evolution in the linear regime leaves the Gaussian nature of the field unchanged. However, on small scales where nonlinear effects are significant, we see both an increase in power in the 2-point measurements (relative to linear evolution) and the density field becomes distinctly non-Gaussian, to the point where it is better described by a log-normal distribution (e.g. \citealt{Colombi94}, \citealt{Clerkin2016Lognormality}). It is thus important to consider methods to extract information beyond the standard 2-point weak lensing statistics.

One possible approach is to user higher-order statistics, such as the 3-point correlation function, to study the non-Gaussianity of the density field (e.g. \citealt{Bernardeau2002Nonlinear3ptReview}). Alternatively, the 1-point distribution of the density field contains information on nonlinear growth and non-Gaussianity. \citealt{Yang2011ComologicalPeaks} discuss the information in peak counts at both the high end $> 3.5 \sigma$ and medium peaks $0.5-1.5 \sigma$ (relative to the mean density) and their origins from dark matter haloes, and find that the peak counts contain non-Gaussian information that is complementary to the power spectrum (see \citealt{Kacprzak16shearpeaks} for a recent measurement). Moments of the 1-point probability density function (PDF) can be combined with the peak count information in order to extract additional information \citep{Petri16}.

The goal of this paper is to characterize the additional information contained in the 1-point PDF of the weak-lensing convergence field, which we refer to as the `convergence PDF' from here on. We focus on the convergence field, which can be reconstructed from weak lensing observations and reflects the projected matter density between the observer and source galaxies. In order to properly determine the information contained in this field, an ideal analysis would consist of running Markov chain Monte-Carlo with a full simulation of the density field to model the likelihood of the convergence PDF at each point in parameter space. However, for an initial analysis we employ the Fisher information framework to analyze the first order Gaussian approximation of the likelihood function near the fiducial maximum and thus estimate the cosmological information content.

Our analysis relies on a large number of simulated weak-lensing convergence fields, with realistic noise properties. These simulations allow us to determine both the cosmological dependence of our measurement and the expected survey-to-survey covariance of this measurement, and thereby determine the Fisher matrix. We measure both the convergence PDF and the power spectrum of the `observed' convergence field in each simulation, allowing us to self-consistently compare the cosmological information content of each type of measurement.

We begin in Section \ref{sec:Simulation Framework} by describing the set of simulations we use and how we create a weak lensing convergence map from the simulation outputs. Section \ref{sec:Measurements} describes how we measure the convergence PDF and the power spectrum of these maps. In Section \ref{sec:Fisher Information} we discuss the Fisher information in these measurements, and finally we discuss the results of the analysis in Section \ref{sec:Results}. Throughout, we use a flat $\Lambda$CDM cosmology with $\Omega_m = 0.25$, $\Omega_b = 0.044$, $\sigma_8 = 0.8$, $h = 0.7$, and $n_s = 0.95$ as our fiducial cosmology. This matches that of the MICE Grand Challenge simulation \citep{Fosalba2015MICEGC}, a single high-resolution cosmological simulation to which we will compare throughout.

\section{Simulated Measurement Framework}

\label{sec:Simulation Framework}

We characterize the information content of the convergence PDF based on mock measurements. The first step in this process is to simulate the cosmological density field of the Universe. In Section \ref{subsec:Density Field Simulations} we describe the particular choice of tool we used in simulating the density field. From these mock density fields we can then generate the weak lensing convergence field as described in Section \ref{subsec:Projected Weak Lensing}. Finally, we include the effects of intrinsic shape noise to the projected convergence field as described in Section \ref{subsec:Shape Noise}

\subsection{Simulations of the Density Field}
\label{subsec:Density Field Simulations}

In order to relate measurements to cosmological quantities, we need a method for predicting how the structure of the Universe depends on cosmological parameters. In the nonlinear regime, the standard method for making these predictions is through the use of a cosmological N-body simulation (e.g.\ {\sc GADGET-2}; \citealt{Springel2005GADGET2}). Modern N-body codes evolve a set of particles from initial conditions using effective gravitational forces in a comoving reference frame. Calculating forces between every pair of particles in a large simulation volume is impractical, so modern techniques use approximations involving long-range force calculations through Fast Fourier transforms (also known as particle-mesh methods), while falling back on direct pairwise force evaluations for nearby particles. However, such methods still require a large number of time steps and pairwise calculations, resulting in long runtimes on standard computational resources. For instance, the MICE Grand Challenge simulation required 3.1 million CPU hours for a simulation volume of (3072 Mpc $h^{-1})^3$, represented on a grid with $4096^3$ cells using $4096^3$ particles. 

For this reason, alternative simulation techniques that make various approximations in order to speed up simulation have been developed. One alternative to {\sc GADGET}, based on Lagrangian Perturbation Theory, is COLA (COmoving Lagrangian Acceleration) \citep{Tassev2013COLA}. 

On large scales the time integration in N-body codes primarily solves for the linear growth factor. Using too few time steps in N-body simulations produces poor estimates for this growth factor and the corresponding amount of large scale power; COLA solves this problem by directly computing the linear growth factor and evolving the motion of particles in a frame comoving with the assumed large scale structure growth. At the cost of slight inaccuracies on the smallest scales, one can run simulations in COLA using significantly fewer time steps, typically of order 10 for COLA versus 2000 for {\sc GADGET}.

We use the software framework {\sc L-PICOLA} \citep{Howlett2015LPICOLA}, a `past-lightcone-enabled' COLA implementation, in order to generate simulated density fields. The output generated by {\sc L-PICOLA} includes a particle catalog that details the positions and velocities of every particle on the past lightcone of the observer, thus allowing for the quick generation of simulated density fields that correctly account for the evolution of structure growth.

In any simulation, the average matter density, $\rho_m$, is represented by $N^3$ particles in an $L^3$ volume, resulting in an individual particle mass $m_p$  of
\begin{equation}
m_p = \rho_m \Big( \frac{L}{N} \Big)^3 .
\end{equation}
In {\sc L-PICOLA}, memory and computational complexity scale roughly as the number of particles in the volume, $N^3$. Pairwise force calculations that might scale as $N^6$ are unnecessary because the particles are projected onto a grid that is Fourier transformed to evaluate forces on all particles. One drawback of the method is that unlike {\sc GADGET}, {\sc L-PICOLA} does not use a tree expansion to calculate forces on small scales of order the grid spacing. This also prevents small scale structure from being properly resolved in the {\sc L-PICOLA} simulation when compared to an equivalent {\sc GADGET} simulation.

However, when accepting this loss of resolution on small scales, the runtime of {\sc L-PICOLA} can be several orders of magnitude faster ($\approx200$), allowing us to run the large number of simulations necessary for proper covariance estimation. In this analysis we limit our measurements to scales large enough ($\ell < 886$) that the inaccuracies at small scales are minor (see section \ref{sec:Measurements} for further discussion). Notably, we include realistic shape noise in our analysis, which we find dominates power spectrum measurements at the scales where we begin to find deviations between {\sc L-PICOLA} and {\sc GADGET} simulations. Thus, small inaccuracies in {\sc L-PICOLA} do not have a significant effect on our analysis, and the speed of the simulations allows us to create enough realizations to properly estimate covariances.

\subsection{Projected Weak Lensing}
\label{subsec:Projected Weak Lensing}

For a given matter field, the projected convergence field for background sources at redshift $z_s$ is calculated as \citep{Bartelmann01}

\begin{equation}
\kappa(\theta) = \int \frac{\rho(\theta, z)}{\Sigma_{\rm crit}(z,z_s)} dz ,
\end{equation}
where $\kappa(\theta)$ is value of the convergence field at position $\theta$ on the sky, and $\rho(\theta, z)$ is the matter density at that position and at redshift $z$. $\Sigma_{\rm crit}(z,z_s)$ is the critical surface density for lenses at redshift $z$ that lenses sources at redshift $z_s$,
\begin{equation}
\Sigma_{\rm crit}(z_l,z_s) = \frac{c^2}{4 \pi G}\frac{D_A(0,z_s)}{D_A(z_l,z_s)D_A(0,z_l)} ,
\end{equation}
where $D_A(z_1,z_2)$ is the angular diameter distance between redshifts $z_1$ and $z_2$.

We estimate $\kappa(\theta)$ above using the discrete distribution of particles provided by our simulations, which we project onto a {\sc HEALPix} map \citep{HEALPix2005}. The convergence field thus becomes a sum over the particles in the simulation with representative lensing weights:
 \begin{equation}
\kappa(\theta) = \sum_p \frac{m_p}{A_{\rm pixel} D_a(0,z_l)^2} \frac{1}{\Sigma_{\rm crit}(z,z_s)} ,
\end{equation}
where the mass of the particles $m_p$ is defined above and $A_{\rm pixel}$ is the area of individual {\sc HEALPix} pixels. For our field we assume a fixed source redshift of the background galaxies. Our setup is thus idealized, but that does not affect our main conclusions.

The shot noise contribution to the projected convergence for each pixel due to discrete particles representing the underlying continuous density field is then
\begin{equation}
\sigma_{\kappa,shot}^2(\theta) = \sum_p \Big( \frac{m_p}{A_{\rm pixel} D_a(0,z_l)^2} \frac{1}{\Sigma_{\rm crit}(z,z_s)} \Big)^2 .
\end{equation}

Averaging this quantity over the field gives us a properly weighted shot noise $\langle \sigma_{\kappa,shot}^2 \rangle$ due to the discrete sampling of the simulated density field.

\subsection{Shape Noise}
\label{subsec:Shape Noise}

With projected mass binned on to a {\sc HEALPix} grid, the next step is to add realistic shape noise. The noise on the convergence field in a realistic survey depends on a number of factors, including (but not necessarily limited to): intrinsic shape noise (scatter in the true shapes of galaxies), background galaxy density, the distribution of redshifts of background galaxies, typical signal-to-noise ratio per galaxy, and variations of such quantities across the survey field. In addition, the method used to determine the convergence as reconstructed from a shear catalog also affects the noise properties of a measured convergence map. For our purposes we will assume Kaiser-Squires inversion \citep{KaiserSquires1993} because it is straightforward to calculate and the noise properties of the resulting maps are well understood, since an observed field can be thought of as the linear sum of the intrinsic projected convergence and a noise realization. 

%[ Practical estimator noise properties: Kaiser Squires 1993, section 2.2 ]

\cite{KaiserSquires1993} gives the variance on the projected convergence map estimated from weak lensing shapes as
\begin{equation}
\langle \sigma_\kappa^2 \rangle = \langle e^2 \rangle \int { d^2 k \frac{ T^2(k) } { 8 \pi^2 n_g } } ,
\end{equation}
where $\langle e^2 \rangle$ is the shape noise for the source galaxies,  $n_g$ is the background galaxy density, and $T(k)$ is the transfer function corresponding to the filter applied to the convergence field.

For {\sc HEALPix} pixels we can approximate the filter as a rectangular tophat, resulting in a transfer function 
\begin{equation}
T(k) = \sinc(k_x \Delta \theta_x) \sinc(k_y \Delta \theta_y).
\end{equation}
The result is that for pixels with dimensions $\Delta \theta_x, \Delta \theta_y$, the per-pixel variance due to shape noise is
\begin{equation}
\langle \sigma_\kappa^2 \rangle =  \frac{ \langle e^2 \rangle } { 2 n_g \Delta \theta_x \Delta \theta_y } .
\end{equation}

However, in our case we already have variance on our projected convergence field of the same form due to shot noise in the simulations $\langle \sigma_{\kappa,shot}^2 \rangle$. We can effectively mask this shot noise, an artifact of the low mass resolution of our simulations, by adding a smaller amount of noise to the convergence field in order to end up at the final desired shape noise for the survey.

\section{Measurements}
\label{sec:Measurements}

One {\sc L-PICOLA} simulation provides full-sky coverage, equivalent to multiple surveys worth of observations. For our fiducial cosmology, we ran $200$ {\sc L-PICOLA} simulations. The fiducial survey size we study is 859 square degrees, corresponding roughly to a Year 1 analysis of the Dark Energy Survey. Each full-sky simulation produces $48$ non-overlapping mock realizations (`mocks') of such a survey, resulting in $9600$ total mocks to estimate our signal covariance matrices. We rotate each mock to the same location on the sky in {\sc HEALPix} to perform each measurement. This does not matter when calculating quantities such as the convergence PDF, but the power spectrum estimation using {\sc HEALPix} can have minor variations with location on the sky due to approximations in the pixel window function for $N_{side}>128$.

For each mock convergence field, we subtract the mean convergence across the individual mock (thus making them `mean-subtracted'). This is necessary due to the manner in which weak lensing probes the projected convergence. Galaxies located inside the survey footprint can only capture relative differences in the convergence of that area; information about whether the survey is sitting in an overdense or underdense region relative to the rest of the sky can only be constrained by galaxies outside the survey footprint.

\subsection{Power Spectrum}

The number of pixels in a full sky {\sc HEALPix} map is $12 N_{side}^2$, while the number of independent spherical harmonics $Y_{lm}$ up to a maximum scale $\ell_{max}$ is $(\ell_{max}+1)^2$. Equating the two in order to produce a comparison using an equal number of modes in each basis, we believe that a fair comparison would be to compare power spectrum scales $\ell < \sqrt{12} N_{side}$.

From an individual mean-subtracted mock convergence field, we use {\sc HEALPix} to calculate the power spectrum through the {\sc ANAFAST} method. In {\sc HEALPix} the spherical harmonics are discretized into a linearly independent system up to $\ell < 3 N_{side}$. However, accuracy above $\ell > 2 N_{side}$ is limited without significant iterative backward and forward harmonic transforms of the map and can also suffer from aliasing unless the signal is band-width limited.

Based on the accuracy of {\sc L-PICOLA}, we chose to use a scale of $N_{side}=256$ in which to bin the convergence PDF. An equivalent number of basis modes in spherical harmonics corresponds to scales $\ell<886$. At this $N_{side}$ {\sc ANAFAST} only returns modes up to $\ell < 768$, and due to the limitations described above also has inaccuracies at the high end. For this reason, we choose to bin the maps at a higher resolution with $N_{side} = 1024$ for calculating the power spectrum through {\sc ANAFAST} and then limit the analysis to the desired modes $\ell<886$.

\begin{figure}
\centering
\includegraphics[width=0.5\textwidth]{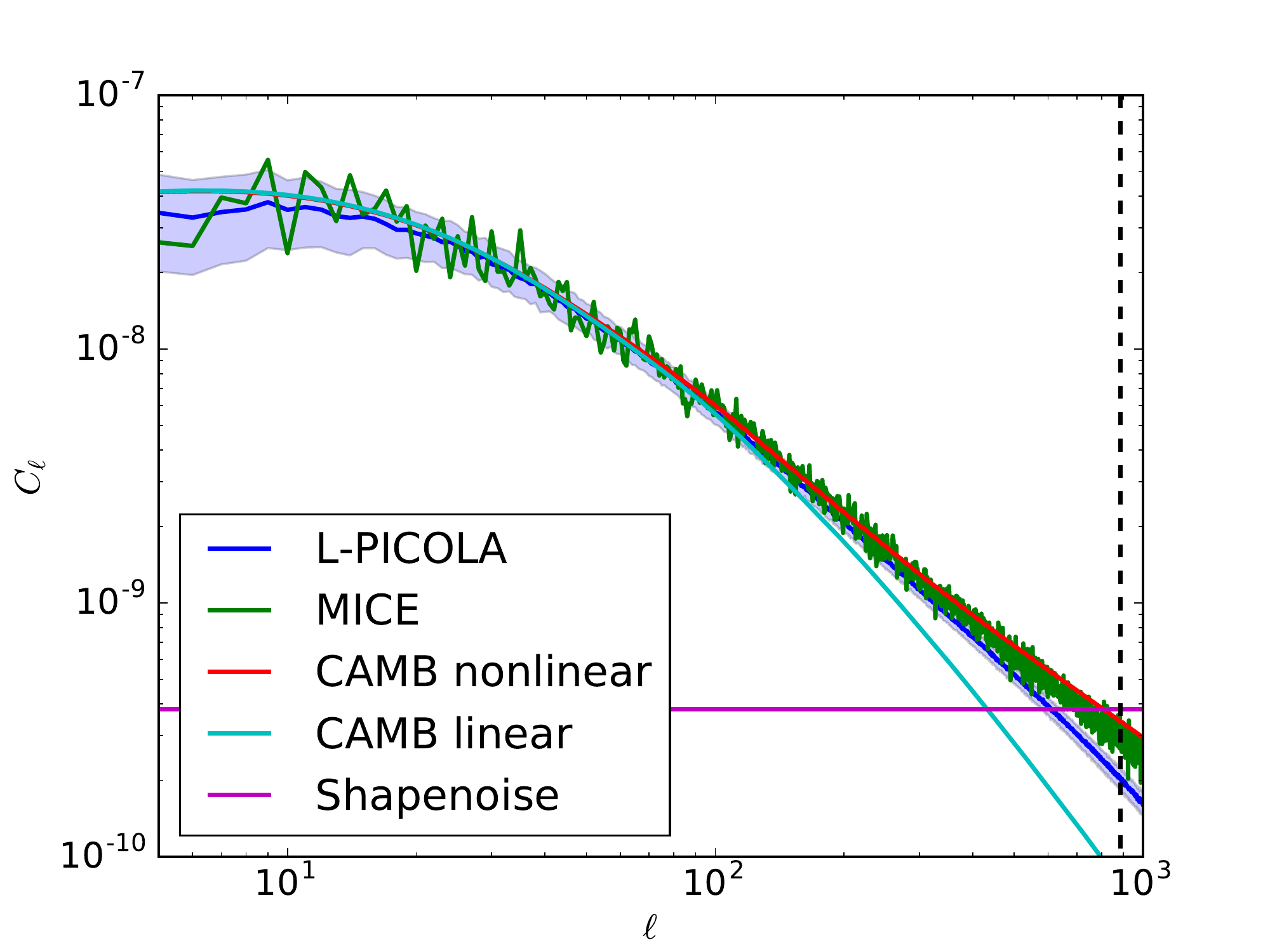}
\caption{\label{fig:PS_CAMB}
	The full-sky power spectrum for different simulation and analytic techniques, all with the fiducial cosmological parameters. {\sc L-PICOLA} is the mean value of the power spectrum from 200 simulations with error bands denoting the standard deviation of the full sky measurement, MICE is the single simulation at the fiducial cosmology run with {\sc GADGET}, and {\sc CAMB} shows both theoretical predictions using {\sc CAMB-sources} with and without nonlinear structure growth. For comparison the power spectrum contribution of the shape noise at $n_{gal}=$10/arcmin$^2$ and $\langle e^2 \rangle = 0.3^2$ is included. The dashed line indicates the maximum $\ell$ used in our analysis.
	}
\end{figure}

Figure \ref{fig:PS_CAMB} shows a comparison of the full sky power spectrum in {\sc L-PICOLA} compared to theoretical values from {\sc CAMB-sources} (\citealt{Lewis2007CambSources,Challinor2011CambSources}) linear and nonlinear {\sc HALOFIT} (\citealt{SmithHalofit,TakahashiHalofit}) power spectrum predictions in addition to the results of an actual {\sc GADGET} run from the MICE grand challenge. We also show the expected level of shape noise for our survey, which uses $n_{gal} = 10/$arcmin$^2$ and shape noise levels of $\langle e^2 \rangle = 0.3^2$. As expected, {\sc L-PICOLA} underestimates the power on small scales. However, it fully captures the linear structure growth and the majority of the nonlinear growth on scales $\ell<886$. 

\subsection{Convergence Probability Density Function}

For the convergence PDF we calculate the histogram of projected convergence values in pixels for each individual survey. Starting from the $N_{side}=1024$ projected convergence maps we downsample to a resolution of $N_{side}=256$ in order to use approximately the same information that is in the power spectrum below $\ell<886$. This results in individual pixels with an area of 188 arcmin$^2$.

\begin{figure}
\centering
\includegraphics[width=0.5\textwidth]{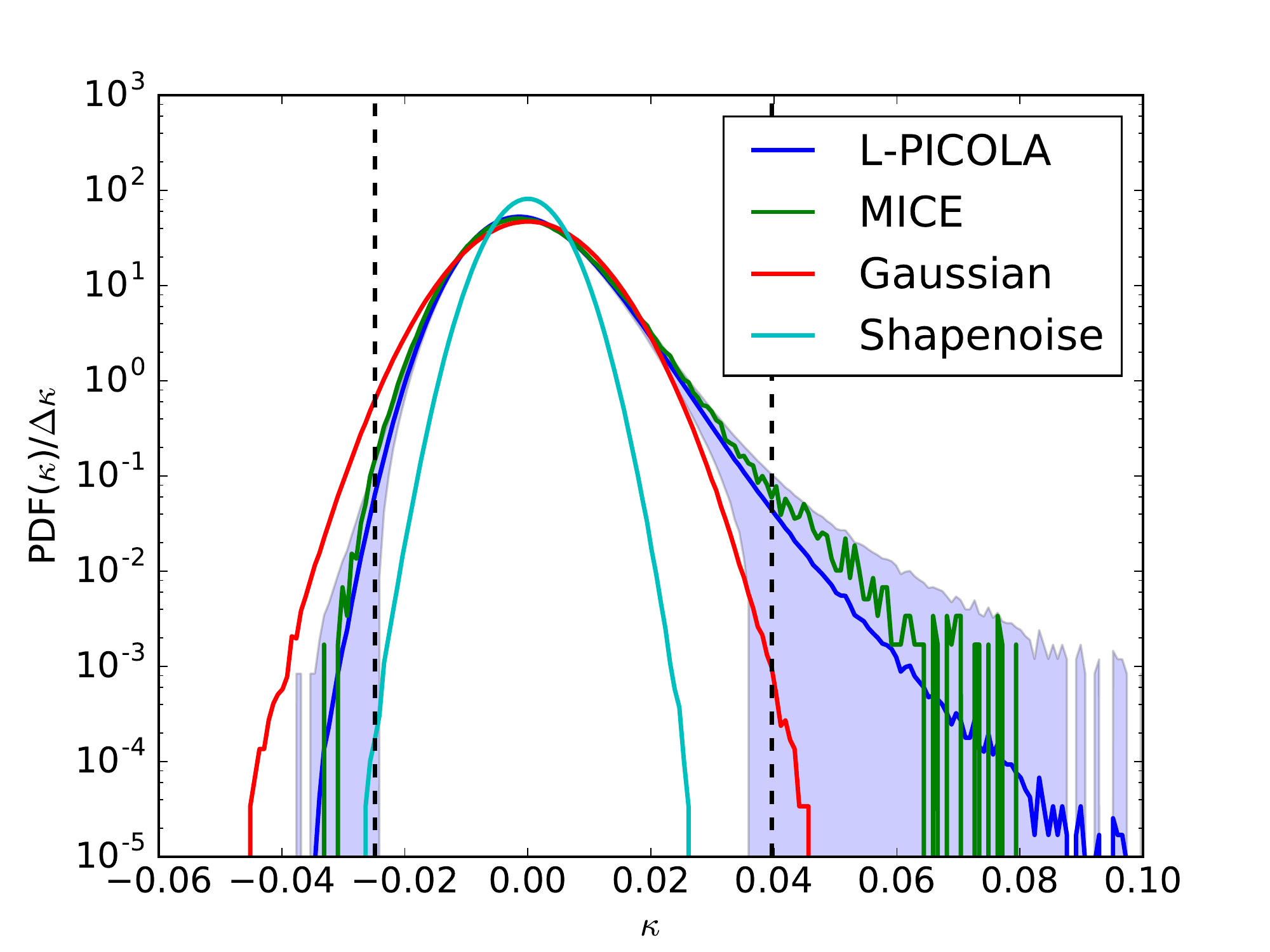}
\caption{\label{fig:PDF_CAMB}
	The convergence PDF at $N_{side}=256$ resolution is shown for both the {\sc L-PICOLA} and MICE simulations as well as the Gaussian result from nonlinear {\sc CAMB}. Shapenoise is included in the convergence PDF distributions and also plotted separately to indicate its effective width. The error bars for the {\sc L-PICOLA} results indicate the expected variance for a 859 deg$^2$ survey containing 16384 pixels. We exclude bins for which the projected convergence value would occur in less than an average of 0.5 pixels per survey, indicated by the region outside the vertical dashed lines. 
	}
\end{figure}

Figure \ref{fig:PDF_CAMB} shows a comparison of the convergence PDF, including shape noise, from the {\sc L-PICOLA} and MICE simulations as well as that from a pure Gaussian field with the equivalent power spectrum from nonlinear {\sc CAMB-sources}. A field with no signal and pure shape noise is plotted for comparison. The cosmological signal clearly deviates from what would be expected in a purely Gaussian field, and this difference is statistically significant even for a single 859 deg$^2$ survey. We also find good agreement between the convergence PDF we estimate with {\sc L-PICOLA} and the MICE simulations at a map resolution of $N_{side}=256$ and below. At a map resolution of $N_{side}=512$ and above the results from {\sc L-PICOLA} and MICE begin to disagree due to inaccuracies in {\sc L-PICOLA}, which is why we limit our analysis to the information available at a map resolution of $N_{side}=256$.

\section{Fisher Forecasts}
\label{sec:Fisher Information}

In this section, we describe how we calculate the Fisher information from our mock results. This Fisher information represents the leading-order Gaussian shape of the likelihood near its maximum. We first present the formalism and then describe how we incorporate systematic uncertainties specific to the measurement of the galaxy shear field into this formalism.

\subsection{Fisher Matrix Formalism}

The Fisher information is calculated as
\begin{equation}
F_{ab} = \frac{dM_i}{dQ_a} \Big ( C ^{-1}_{ij} \Big ) \frac{dM_j}{dQ_b} ,
\end{equation}
where $C_{ij}$ is the covariance matrix of the measurements, and $\frac{dM_i}{dQ_a}$ is the derivative of the measurement signal in bin $i$ with respect to cosmological parameter $Q_a$. The diagonal elements of the inverse of the Fisher matrix $F_{ab}$ indicate the expected constraints on parameters $Q_a$, marginalized over the other parameters. Off-diagonal elements indicate correlations between various cosmological and nuisance parameters.

We need to estimate both the covariance of the signal $C_{ij}$ and the cosmological derivatives of our measurements $\frac{dM_i}{dQ_a}$. The covariances are estimated by performing 200 simulations at the fiducial cosmology, where each simulation yields 48 independent mocks for a total number of 9600 mocks. In general
\begin{equation}
C_{ij} = \sum_{k=0}^{N} \frac{1}{N-1}(X^k_i-\bar{X_i})(X^k_j-\bar{X_j}), 
\end{equation}
where, e.g., $X_i$ can represent a bin in our convergence PDF histogram and $N$ is the number of mocks used.

We estimate the cosmological derivatives of the signal through finite differences of the signal between the fiducial cosmology and cosmologies in which each parameter of interest is varied individually. For this analysis $\Omega_k=0$ and $n_s= 0.95$ are held fixed while $\Omega_m$, $\sigma_8$, and $H_0$ are varied. We then apply a prior on the final Fisher matrix in order to hold the product $\Omega_m h^2$ (which is tightly constrained by CMB measurements) fixed for the final analysis. The set of cosmological parameters we considered are described in Table \ref{tab:cosmo}, which also includes the finite differences by which the cosmological parameters were varied. For each parameter varied, we performed 50 simulations at the new cosmology, yielding 2400 mocks to calculate the mean signal.

\begin{table*}
	\centering
	\begin{tabular}{c c c}
		
		parameter & fiducial value & delta  \\
		\hline
		$\Omega_m$ & 0.25 & 0.03 \\
		$\sigma_8$ & 0.80 & 0.08 \\
		$\Omega_k$ & 0.0 & - \\
		$h$ & 0.70 & 0.07 \\
		$w$ & -1.0 & - \\
		m & 0.0 & 0.1 \\
		c & 0.0 & 0.1 \\
		$\Omega_b$ & 0.044 & - \\
		$n_s$ & 0.95 & - 
		
	\end{tabular}
	\caption{\label{tab:cosmo}
		Cosmological parameter values for finite difference model evaluation. Fiducial simulation values are listed along with the finite difference used to evaluate the Fisher information content of the cosmological signals.
		}
\end{table*}

Estimating the covariance from a limited number of simulations can induce a systematic bias in the inverse covariance matrix. \cite{Dodelson2013Covariance} and \cite{Percival2014Covariance} discuss this error in covariance estimation, the impact on the Fisher information, and the correction factors required. In addition, we found that an additional correction factor was necessary to correct biases on the Fisher information due to errors on our estimates of the finite-difference derivatives $\frac{dM_i}{dQ_a}$. These cosmological derivatives are estimated from a finite number of simulations, which will contain fluctuations from the true cosmological signal consistent with the covariance of the measurements. Although these fluctuations are in general unbiased for $\frac{dM_i}{dQ_a}$, nonlinear combinations of the products of these cosmological derivatives, such as those in the Fisher matrix, can in general lead to a bias. 

In the appendix we derive the required correction factor:

\begin{equation}
F_{ab} \approx \widehat{F_{ab}} - \frac{N_{bin}}{N_f \Delta Q^a \Delta Q^b} - \frac{N_{bin}}{N_a \Delta Q^a \Delta Q^b} \delta_{ab} ,
\end{equation}
where the first term is the Fisher matrix estimated from simulations, the second term is a correction factor resulting from a bias due to covariance of the estimate of the fiducial cosmological signal, and the last term is a bias resulting from covariance of the signal in the varied cosmologies. $N_{bin}$ is the number of bins in the cosmological signal, $N_{f}$ is the number of mocks used for estimating the mean fiducial cosmological signal, $N_{a}$ is the number of mocks used to calculate the mean signal at the differenced cosmology, and $\Delta Q^a$ is the amount by which parameter $Q_a$ was varied.

\begin{figure}
\centering
\includegraphics[width=0.5\textwidth]{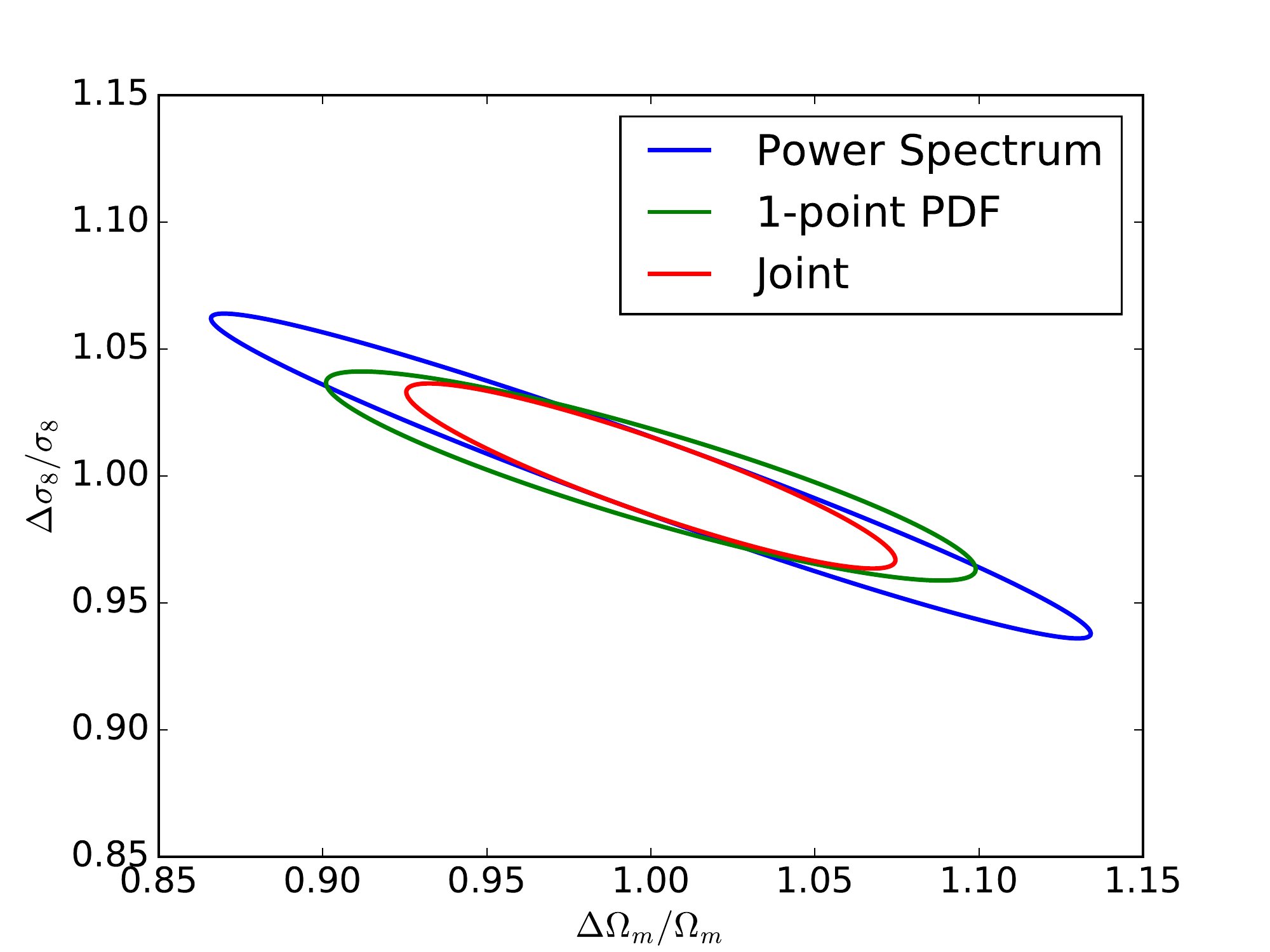}
\includegraphics[width=0.5\textwidth]{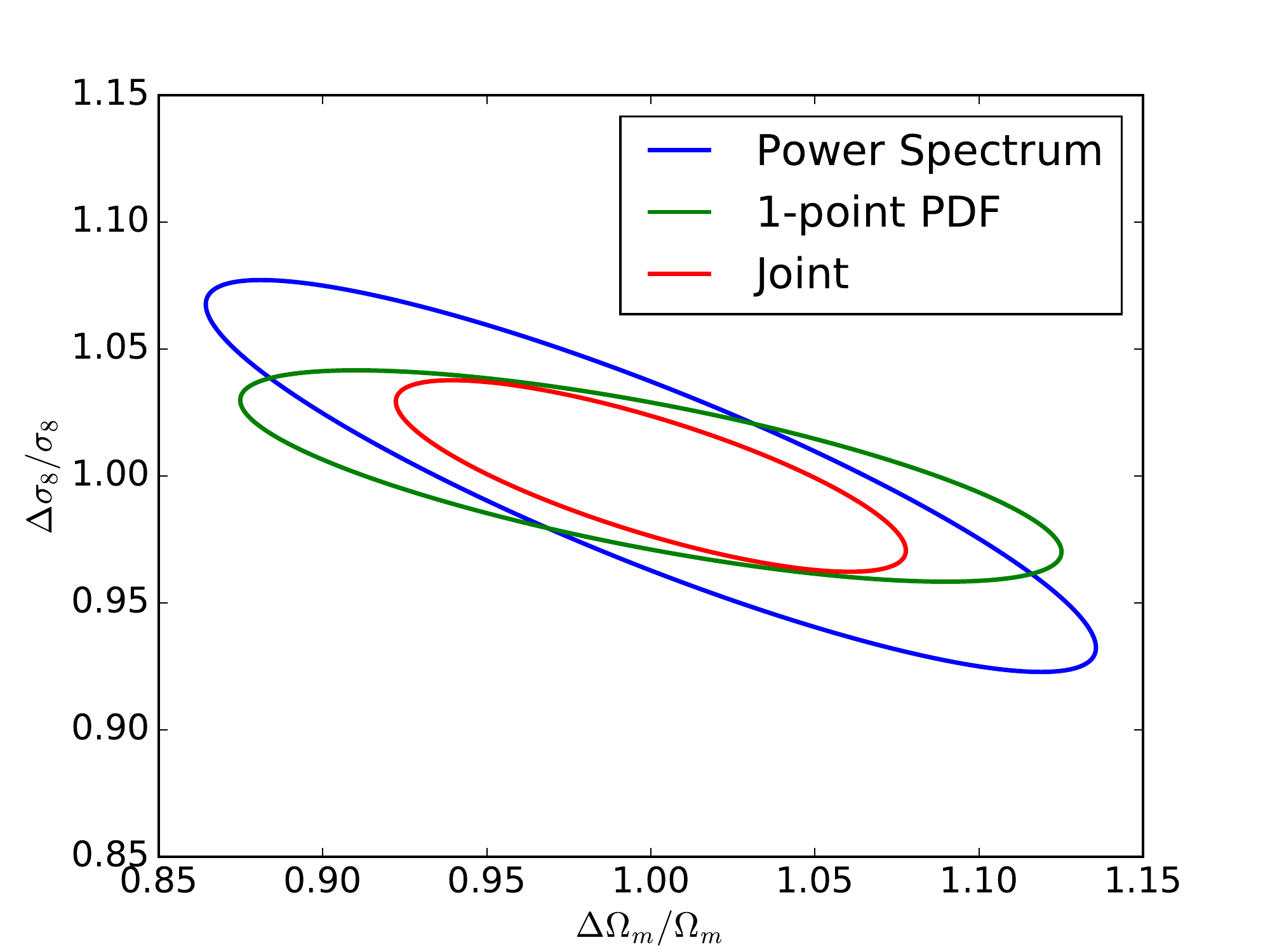}
\caption{\label{fig:FISHER_COMPARE}
	Fisher constraints are shown for a 859\ deg$^2$ survey. The power spectrum includes bins $\ell < 886$ and the convergence PDF is for 188\ arcmin$^2$ pixels  within the 859\ deg$^2$ survey (16384 total pixels). Joint constraints use the full covariance information between power spectrum and convergence PDF bins. The top figure includes no measurement systematics while the bottom figure includes unknown multiplicative and additive shear biases with Gaussian priors of $\Delta m = 0.05$ and $\Delta c = 0.0123$; see Eq. \ref{eq:sys}.
	}
\end{figure}

\subsection{Measurement Systematics}
\label{subsec:Measurement Systematics}

There are two main sources of systematic uncertainty that we account for in our Fisher analysis: additive and multiplicative biases on the shear signal \footnote{In practice, we will also encounter errors in the redshift determination of the sources, but to first order they act as multiplicative errors on the lensing signal}. We define these contributions as\begin{equation}
\widehat{\kappa}=(1+m)[\kappa + e(1 + c)] ,
\label{eq:sys}
\end{equation}
where our estimator for the projected convergence $\widehat{\kappa}$ is related to the true projected convergence $\kappa$ and shape noise $e$ by the additional parameters $m$ and $c$, respectively the multiplicative and additive shear biases. We leave for future work the impact of other systematics such as baryonic physics and intrinsic alignments (e.g.\ \cite{Eifler15, Blazek15}), although we note that our conservative choice of minimum scale should minimize the effect of baryons. 

Considering these biases as part of our parameter set $Q = \{\Omega_m, \sigma_8, h, m, c\}$, we can determine how a lack of knowledge of these systematics can affect the constraints on both $\Omega_m$ and $\sigma_8$. \cite{Jarvis2016DESSV} discusses the tolerances for these bias parameters measured from the Dark Energy Survey Science Verification survey area. They find that the tolerances measured from the survey can be estimated to be at most $\Delta c < 0.0123$ for the additive systematic and $\Delta m < 0.05$ for the multiplicative bias. These are estimated from a survey similar to our fiducial one but with a smaller area, so we adopt their recommended additive and multiplicative systematic estimates as Gaussian priors in our analysis. In practice one would prefer to reduce these systematics for the larger survey area we consider here, but these priors represent the current level of shear estimation systematics as a pessimistic view of the future in case they are unable to be reduced further.

\section{Results}
\label{sec:Results}

\begin{figure}
	\centering
	\includegraphics[width=0.5\textwidth]{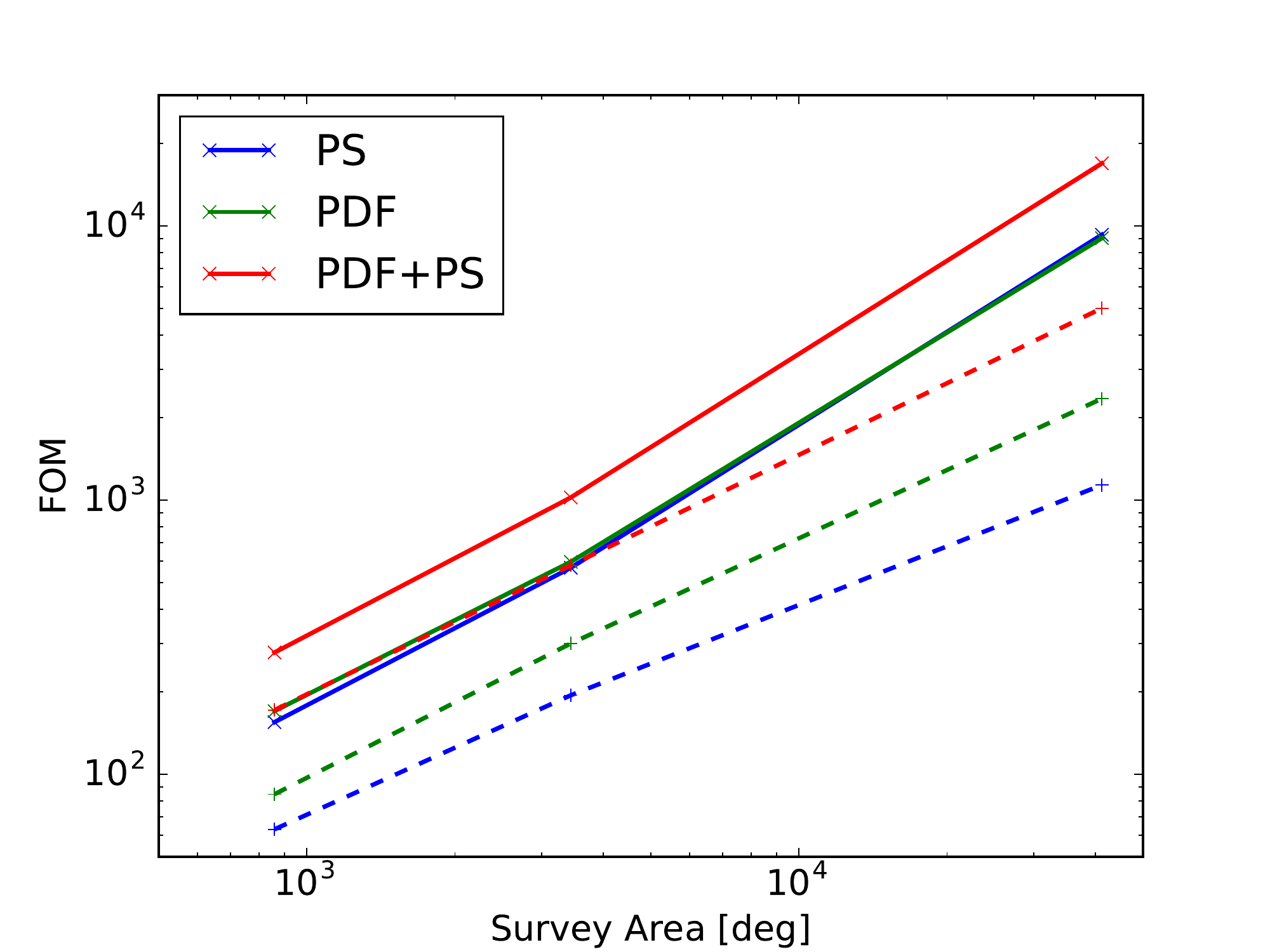}
	\caption{	\label{fig:FISHER_SCALING_FOM}
		The solid lines show how the figure of merit scales with survey area, when no systematics are included. The dashed line show the scaling when including additive and multiplicative systematics with priors of $\Delta m = 0.05$ and $\Delta c = 0.0123$. The symbols indicate the actual survey areas for which we calculated the Fisher information to obtain the FOM.		
	}
\end{figure}

Figure \ref{fig:FISHER_COMPARE} shows the Fisher constraints for a 859\ deg$^2$ survey with and without accounting for shear systematics. These parameter constraints are also listed in Table \ref{tab:param constraints}. In order to examine the primary degeneracy directions, we can also consider rotation of these cosmology parameters to a new set of orthogonal parameters:

\begin{multline}
Q_1 = \frac{\alpha}{\sqrt{1+\alpha^2}}\frac{\Omega_m}{\Omega_m^{fid}} + \frac{1}{\sqrt{1+\alpha^2}}\frac{\sigma_8}{\sigma_8^{fid}} ,
\end{multline}
and
\begin{multline}
Q_2 = \frac{1}{\sqrt{1+\alpha^2}}\frac{\Omega_m}{\Omega_m^{fid}} - \frac{\alpha}{\sqrt{1+\alpha^2}}\frac{\sigma_8}{\sigma_8^{fid}} .
\end{multline}

For the correctly chosen value of $\alpha$, these parameters correspond to the minor and major axes of constraint ellipses (i.e.\ the best- and worst-constrained parameter combinations) and also represent the basis in which the Fisher matrix is diagonal. Note that this parameter rotation preserves the overall area in the 1-sigma uncertainty region. \cite{Jain1997Degeneracy} discuss the expected cosmological degeneracies of $\Omega_m$ and $\sigma_8$ for power spectrum measurements, finding that the power spectrum approximately measures the product $\sigma_8 \Omega_m^\alpha$ for $\alpha \approx 0.5$, depending on cosmological model, source lensing redshift, and scale. To first order the parameter $\sigma_8 \Omega_m^\alpha$ is equivalent to our parameter $Q_1$.

In Table \ref{tab:param constraints} we list the constraints at the fiducial value of $\alpha=0.5$, in addition to the best-fit value of $\alpha$ that removes any correlation between $Q_1$ and $Q_2$.

\begin{table*}
\centering
\begin{tabular}{c c c c c c c c c c}
\hline
\hline
analysis & $\Delta\Omega_m / \Omega_m$ & $\Delta\sigma_8 / \sigma_8$ & $\Delta Q_1(\alpha=0.5)$ & $\Delta Q_2(\alpha=0.5)$ & $\alpha$ & $\Delta Q_1$ & $\Delta Q_2$ & $\pi \Delta Q_1 \Delta Q_2$ & FOM\\
\hline
no sys PS & 0.134 & 0.064 & 0.015 & 0.148 & 0.468 & 0.014 & 0.148 & 0.007 & 153.415 \\
no sys PDF & 0.099 & 0.041 & 0.020 & 0.105 & 0.383 & 0.017 & 0.106 & 0.006 & 172.631 \\
no sys PDF+PS & 0.075 & 0.036 & 0.014 & 0.082 & 0.459 & 0.014 & 0.082 & 0.004 & 277.687 \\
sys PS & 0.136 & 0.077 & 0.033 & 0.152 & 0.530 & 0.033 & 0.153 & 0.016 & 63.042 \\
sys PDF & 0.125 & 0.042 & 0.039 & 0.126 & 0.251 & 0.028 & 0.129 & 0.011 & 87.764 \\
sys PDF+PS & 0.078 & 0.038 & 0.023 & 0.083 & 0.411 & 0.022 & 0.084 & 0.006 & 172.886 \\
\hline
\end{tabular}
\caption{\label{tab:param constraints}
	Fisher information marginalized parameter constraints for various combinations of measurements with and without systematic errors. We include estimates of errors in the rotated parameter plane $Q_1$ and $Q_2$ for the theoretically-motivated rotation $\alpha=0.5$, in addition to calculating the optimal $\alpha$ to diagonalize the Fisher information matrix. We present the $Q_1$ and $Q_2$ for this optimal $\alpha$ as well as the area of the Fisher constraints and the figure of merit which is the inverse of the Fisher constraint area.
	}
\end{table*}

We find that without including shear systematics, the power spectrum provides slightly better constraints along the best-constrained axis than the convergence PDF. However, along the least constrained axis, the convergence PDF provides significantly more constraining power. As a result, the convergence PDF provides tighter constraints on either $\Omega_m$ or $\sigma_8$ alone when marginalizing over the other. We also define the figure of merit (FOM) as the inverse area of the 1-sigma uncertainty region in the $\Omega_m / \Omega_m^{fid} , \sigma_8 / \sigma_8^{fid}$-plane, finding a greater FOM for the convergence PDF alone compared to the power spectrum. Combining the measurements using the full joint covariance results in further improvements over either method alone.

Similarly, we compare the constraints when including additive and multiplicative shear biases and find qualitatively similar results to the case with no systematics. The main added benefit from the convergence PDF information appears to be in helping break the large degeneracy between $\Omega_m$ and $\sigma_8$ that is inherent to using only the power spectrum information. 

In Fig. \ref{fig:FISHER_SCALING_FOM} we show how the FOM scales with survey area, for the case where our priors on the level of systematics are held constant (solid curves) and no systematics (dashed curves). Both measurements are affected by a degeneracy between the multiplicative bias and $\sigma_8$. Of particular note is that the information in the convergence PDF is less affected by this degeneracy, indicating that it can provide a substantial increase in the amount of cosmological information available even in the case that the multiplicative bias cannot be reduced in future large scale surveys. In this analysis, the assumed level of multiplicative bias was fixed at the level recently determined in the analysis of Dark Energy Survey data \citep{Jarvis2016DESSV}. The relative benefit of the convergence PDF information will depend on the precise level of systematic uncertainty, but it provides substantial improvement over the power spectrum alone in all cases.

\section{Discussion and Conclusions}

In this paper we study the amount of information in the PDF of the convergence of the weak lensing shear field. We find that in the $\Omega_m - \sigma_8$ space, PDF information improves the size of expected constraints by a factor of two. The gains from the convergence PDF are even greater when multiplicative shear biases dominate compared to the overall statistical power. The improved constraining power mostly results from tightening the least constrained axis of the Fisher contours in the $\Omega_m - \sigma_8$ space, as the power spectrum has a slight advantage in the best constrained combination of the parameters. We also find that the direction of the main degeneracy between $\Omega_m$ and  $\sigma_8$ is similar between both the convergence PDF and the power spectrum, although combining the two measurements still provides a significant improvement on the Fisher constraints.

We compare our results to other studies that have considered the information content of the shear field beyond that of 2-point statistics. \cite{Kacprzak16shearpeaks} compared constraints from shear peak statistics on aperture mass maps in the Dark Energy Survey Science Verification data for peaks with signal-to-noise between $0 < S/N < 4$, and found similar constraints in the $\Omega_m - \sigma_8$ space between the 2-point non-tomographic power spectrum and shear peak statistics. We note that when we limit our convergence PDF to values of $\kappa > 0$, the figure of merit drops by around $20 \%$ and yields similarly comparable constraints to the power spectrum. For this reason we believe it is important to note that convergence values below zero provide valuable information that is not available in the peaks.

\cite{Liu2015PeakCounts} compares weak lensing peak counts to the power spectrum for CFHTLenS observations, and find that peak counts provide comparable information to the power spectrum. Combining the two, the size of the constraints in the $\Omega_m$-$\sigma_8$ plane is reduced by roughly a factor of 2. This is similar to the factor of $\approx 2$ improvement we find for the case with no systematics. \cite{Petri16} forecast parameter constraints in the ($\Omega_m$, $w$, $\sigma_8$) space and similarly find that adding non-Gaussian information from peak counts and moments of the convergence distribution adds significant information even when compared to tomographic power spectrum results.

Overall, the non-Gaussian features of the convergence map provide significant information beyond the power spectrum and should be an important component for an analysis of cosmological surveys. Even without systematics the addition of the convergence PDF significantly improves the figure of merit from the power spectrum alone. This additional information will be especially useful in a scenario where measurement systematics cannot be reduced in future large area survey, indicating the convergence PDF may become a crucial component for robust cosmological inference from weak lensing.

%\section*{Acknowledgements}
%JB and AJR are supported through CCAPP Fellowships...

\bibliography{paper}

\newpage

\clearpage
\newpage

\section{Appendix}

In this appendix, we derive a correction factor required for unbiased Fisher information when using a finite number of simulations to estimate the response of the signal to a change in underlying cosmology.

Let us define our measurement $M_i^x(Q)$,
where $i$ signifies the bins in our measurement and $x$ is an index representing the possible measurement draw we obtained from cosmology $Q$. The true covariances for our measurement at cosmology $Q$ are
\begin{multline}
C_{ij}(Q) = \Big  \langle \Big (M^x_i(Q) - \overline{M_i}(Q)\Big ) \Big(M_j^x(Q) - \overline{M_j}(Q)\Big) \Big \rangle_x \\
= \Big  \langle M^x_i(Q) M_j^x(Q) \Big \rangle_x - \Big ( \overline{M_i}(Q) \overline{M_j}(Q) \Big ) ,
\end{multline}
where the mean value is
\begin{equation}
\overline{M_i}(Q) = \Big \langle M^x_i(Q) \Big \rangle_x ,
\end{equation}
averaging over all possible measurement draws $x$. The Fisher information that tells us about the Gaussian likelihood distribution around the maximum likelihood value of cosmological parameters $Q^f = \{Q^f_1, Q^f_2, ...\} = \{Q^f_a\}$ is

\begin{equation}
F_{ab} = \sum_{ij} \frac{d\overline{M_i}}{dQ_a}\Big(C^{-1}\Big)_{ij}\frac{d\overline{M_j}}{dQ_b} .
\end{equation}

If we consider using finite differences from a limited number of simulations to estimate $\frac{d\overline{M_i}}{dQ_a}(Q^{f})$ we have
\begin{equation}
\widehat{M_i}(Q^f) = \frac{1}{N_f}\sum_{x}^{N_f} M_i^x(Q^f) ,
\end{equation}
and 
\begin{equation}
\widehat{M_i}(Q^f+\Delta Q^a) = \frac{1}{N_a}\sum_{y}^{N_a} M_i^y(Q^f+\Delta Q^a) ,
\end{equation}
where we have $N_f$ realizations of the measurement from simulations at the fiducial cosmology $Q^f$ and $N_a$ realizations of the measurement at each differenced cosmology $Q^f + \Delta Q^a$. Then the finite difference estimates of the cosmological derivatives for our measurements are
\begin{equation}
\widehat{\frac{d{M_i}}{dQ_a}} = \frac{ \widehat{M_i}(Q^f+\Delta Q^a) - \widehat{M_i}(Q^f)} {\Delta Q^a} .
\end{equation}

These estimates of the cosmological derivatives necessarily contain errors due the limited number of simulations used to reduce covariance noise. For a single cosmological derivative these errors are unbiased around the true mean value. However, when considering higher order correlations, such as for the Fisher information, the product of two of these cosmological derivative estimates can result in a bias. Let us consider the product
\begin{multline}
\Big \langle \widehat{\frac{d{M_i}}{dQ_a}} \widehat{\frac{d{M_j}}{dQ_b}} \Big \rangle = \\
\Big \langle \frac{ \widehat{M_i}(Q^f+\Delta Q^a) - \widehat{M_i}(Q^f)} {\Delta Q^a} \frac{ \widehat{M_j}(Q^f+\Delta Q^b) - \widehat{M_j}(Q^f)} {\Delta Q^b} \Big \rangle \\
= \frac{1}{\Delta Q^a \Delta Q^b}\Big \langle \Big ( \widehat{M_i}(Q^f+\Delta Q^a) - \widehat{M_i}(Q^f) \Big ) \\ \Big ( \widehat{M_j}(Q^f+\Delta Q^b) - \widehat{M_j}(Q^f) \Big ) \Big \rangle \\
= \frac{1}{\Delta Q^a \Delta Q^b} \Big ( \Big \langle \widehat{M_i}(Q^f+\Delta Q^a) \widehat{M_j}(Q^f+\Delta Q^b) \Big \rangle \\ + \Big \langle \widehat{M_i}(Q^f) \widehat{M_j}(Q^f) \Big \rangle - \Big \langle \widehat{M_i}(Q^f+\Delta Q^a) \widehat{M_j}(Q^f) \Big \rangle  \\ - \Big \langle \widehat{M_i}(Q^f) \widehat{M_j}(Q^f+\Delta Q^b) \Big \rangle \Big ) .
\end{multline}
The individual terms are
\begin{multline}
\Big \langle \widehat{M_i}(Q^f) \widehat{M_j}(Q^f) \Big \rangle = \Big \langle \frac{1}{N_f}\sum_{x}^{N_f} M_i^x(Q^f) \frac{1}{N_f}\sum_{y}^{N_f} M_i^y(Q^f) \Big \rangle \\
= \frac{1}{N_f}\sum_{x}^{N_f} \frac{1}{N_f}\sum_{y}^{N_f} \Big \langle  M_i^x(Q^f)  M_i^y(Q^f) \Big \rangle .
\end{multline}
Since these are drawn from the same set of samples, when $x = y$ we are averaging the same sample with itself yielding a bias
\begin{multline}
\Big \langle  M_i^x(Q^f)  M_j^y(Q^f) \Big \rangle = \overline{M_i}(Q^f) \overline{M_j }(Q^f) + \delta_{xy} C_{ij}(Q^f) ,
\end{multline}
and thus
\begin{multline}
\Big \langle \widehat{M_i}(Q^f) \widehat{M_j}(Q^f) \Big \rangle = \frac{1}{N_f}\sum_{x}^{N_f} \frac{1}{N_f}\sum_{y}^{N_f} \Big ( \overline{M_i}(Q^f) \overline{M_j }(Q^f) \\ + \delta_{xy} C_{ij}(Q^f) \Big ) \\
	= \overline{M_i}(Q^f) \overline{M_j }(Q^f) + \frac{C_{ij}(Q^f)}{N_f} ,
\end{multline}
Similarly, for the cosmological derivatives
\begin{multline}
\Big \langle \widehat{M_i}(Q^f+\Delta Q^a) \widehat{M_j}(Q^f + \Delta Q^b) \Big \rangle = \\ \overline{M_i}(Q^f + \Delta Q^a) \overline{M_j}(Q^f + \Delta Q^b) + \delta_{ab} \frac{C_{ij}(Q^f+\Delta Q^a)}{N_a} \\
\approx \overline{M_i}(Q^f + \Delta Q^a) \overline{M_j}(Q^f + \Delta Q^b) + \delta_{ab} \frac{C_{ij}(Q^f)}{N_a} ,
\end{multline}
where the $\delta_{ab}$ results from the fact that the different cosmological derivatives come from independent sets of samples. For a similar reason
\begin{multline}
\Big \langle \widehat{M_i}(Q^f+\Delta Q^a) \widehat{M_j}(Q^f) \Big \rangle = \overline{M_i}(Q^f + \Delta Q^a) \overline{M_j}(Q^f) ,
\end{multline}
and
\begin{multline}
\Big \langle \widehat{M_i}(Q^f) \widehat{M_j}(Q^f+\Delta Q^b) \Big \rangle = \overline{M_i}(Q^f ) \overline{M_j}(Q^f + \Delta Q^b) .
\end{multline}

In total, we find
\begin{multline}
\label{eq:bias1}
\Big \langle \widehat{\frac{d{M_i}}{dQ_a}} \widehat{\frac{d{M_j}}{dQ_b}} \Big \rangle \approx \frac{d\overline{M_i}}{dQ_a} \frac{d\overline{M_j}}{dQ_b} + \frac{C_{ij}(Q^f)}{N_f \Delta Q^a \Delta Q^b} + \\
\frac{C_{ij}(Q^f)}{N_a \Delta Q^a \Delta Q^b} \delta_{ab} .
\end{multline}

The first term in Eq.~\ref{eq:bias1} is the true product of the average cosmological derivatives, the second term results as a bias for all products of cosmological derivatives due to error from limited simulation statistics at the fiducial cosmology, and the last term includes a $\delta_{ab}$ and is thus the bias on the product of each cosmological derivative with itself due to limited simulation statistics at the differenced cosmologies. If we consider the impact these errors have on the Fisher information we have
\begin{multline}
\Big \langle \widehat{F_{ab}} \Big \rangle = \Big \langle \sum_{ij} \widehat{\frac{dM_i}{dQ_a}}\Big(C^{-1}\Big)_{ij}\widehat{\frac{dM_j}{dQ_b}} \Big \rangle \\
 = \Big \langle \sum_{ij} \Big(C^{-1}\Big)_{ji} \widehat{\frac{dM_i}{dQ_a}} \widehat{\frac{dM_j}{dQ_b}} \Big \rangle \\
 \approx \sum_{ij} \Big(C^{-1}\Big)_{ji} \Big ( \frac{d\overline{M_i}}{dQ_a} \frac{d\overline{M_j}}{dQ_b} + \frac{C_{ij}}{N_f \Delta Q^a \Delta Q^b} + \frac{C_{ij}}{N_a \Delta Q^a \Delta Q^b} \delta_{ab} \Big ) \\
  \approx F_{ab} + \sum_{ij} \Big(C^{-1}\Big)_{ji} \Big ( \frac{C_{ij}}{N_f \Delta Q^a \Delta Q^b} + \frac{C_{ij}}{N_a \Delta Q^a \Delta Q^b} \delta_{ab} \Big ) \\
  \approx F_{ab} + \frac{N_{bin}}{N_f \Delta Q^a \Delta Q^b} + \frac{N_{bin}}{N_a \Delta Q^a \Delta Q^b} \delta_{ab} ,
\end{multline}
where $N_{bin}$ is the number of bins in the measurement which is equal to the number of rows / columns in the covariance matrix. Correcting for this error, an unbiased estimate of the Fisher matrix due to covariance errors contributing to the model is
\begin{multline}
F_{ab} \approx \widehat{F_{ab}} - \frac{N_{bin}}{N_f \Delta Q^a \Delta Q^b} - \frac{N_{bin}}{N_a \Delta Q^a \Delta Q^b} \delta_{ab} .
\end{multline}
	
\end{document}